# Cold numbers: Superconducting supercomputers and presumptive anomaly

by

Nicola De Liso[1] (Department of Law – Economics Division, University of Salento, via per Monteroni snc, 73100 Lecce, Italy, email: nicola.deliso@unisalento.it)
Giovanni Filatrella (Department of Science and Technology, University of Sannio, 82100 Benevento, Italy, email: filatrella@unisannio.it)
Dimitri Gagliardi (Manchester Institute of Innovation Research, University of Manchester, Oxford Road, M13 9PL, Manchester, UK, email: Dimitri.Gagliardi@manchester.ac.uk)
Claudia Napoli (Department of Physics, University of Salerno, Via Giovanni Paolo II, I, 84084 Fisciano, SA, Italy, email: cnapoli@unisa.it)

## Abstract

In February 2014 *Time* magazine announced to the world that the first quantum computer had been put in use. One key component of this computer is the 'Josephson-junction', a *super*conducting device, based on completely different scientific and technological principles with respect to *semi*conductors. The origin of superconductors dates back to the 1960s, to a large-scale 20-year long IBM project aimed at building ultrafast computers. We present a detailed study of the relationship between Science and Technology making use of the theoretical tools of *presumptive anomaly* and *technological paradigms*: superconductors were developed whilst the semiconductors revolution was in full swing. We adopt a historiographical approach - using a snowballing technique to sift through the relevant literature from various epistemological domains and technical publications - to extract theoretically robust insights from a narrative which concerns great scientific advancements, technological leaps forward and business-driven innovation. The study we present shows how technological advancements, business dynamics and policy intertwine.

---

[1] Corresponding author: nicola.deliso@unisalento.it

## 1. Introduction

In February 2014 the quantum computer revolution was announced to the world through the cover of *Time* magazine which referred to the "infinity machine" produced by the company D-Wave. Scientists, technologists, policy makers and the society at large were particularly impressed by the word 'quantum' associated to 'computing', while some observers reacted to the news with some scepticism (e.g. Shin et al., 2014).

Whether quantum or not, this is a new type of computer whose speed depends on the use of 'Josephson junctions', which – it must be stressed – are *not* based on semiconducting processors, but on a completely different technology which makes use of the property of 'superconductivity'. The latter is a property of certain materials which become superconductors when they are cooled to near-absolute-zero temperatures, i.e. -273 °C.

Early attempts at using *super*conductors in computers date back to the 1960s, and for a while, superconductors emerged as a credible alternative to semiconductors. Superconductors disappeared from the computer industry horizon in the early 1980s (before re-emerging, as an exception, in the 2010s).

In this work we study the emergence and eclipse of the potentially revolutionary technological superconductors' paradigm, which developed along the incumbent semiconductors' paradigm for nearly twenty years.

We believe that this is an important case in which the forces of innovation are at work: the relationships between science and technology, the importance of a Schumpeterian giant firm – IBM –, the role played by technological expectations, governments' attention to a strategic technology, and more.

Our analysis is carried out in an evolutionary theoretical setting, and, specifically, we make use of the concepts of *technological paradigms* and *presumptive anomaly*. We investigate whether a presumptive anomaly gave rise to the search of a new technological paradigm, and whether that search led to a failed innovation. The answer to the first question is "yes", while the answer to the second question is not so clear-cut, and an apt phrase to word the answer is that this is a false-failed innovation[1].

---

[1] The use of these words is due to Wilmoth (2000).

Often scholars focus on cases which can clearly be classified as successful innovations, while only occasionally attention is given to cases of failures (Freeman, 1973; Rothwell *et al*. 1974; Braun, 1992; Freeman and Soete, 1997; Leoncini, 2016).

We propose an interpretation of what happened. Whilst the history of computing is characterised by systematic announcements of revolutionary breakthroughs and new disruptive technologies that rarely materialise or have little impact, the case of superconducting computers deserves explicit attention. IBM was at the centre of the early struggle and invested massive human and financial resources on a long-term project spanning from 1964 to 1983. The explicit aim was that of developing a new technology capable of multiplying up to 100 times computing rates; the financial and human resources invested were consistent with such an ambition, i.e. efforts were not limited to a 'small' laboratory or pilot study.

After forty years in obscurity the superconductors' technology shows tangible signs of life in the computer industry. In 2013 the company D-Wave delivered its first 'infinity machine', which makes use of superconducting Josephson junctions, and in December 2014 the American Intelligence Advanced Research Projects Activity (IARPA) announced that it "has embarked on a multi-year research effort to develop a superconducting computer" (ODNI, 2014: 1). In July 2015, Hypres – a high-tech company – disclosed that it was developing Josephson-junction-based superconducting circuits (Hypres, 2015). These developments find their origin in the first long-run project that began in the 1960s.

This paper is articulated as follows. In section 2 we present theoretical background within which we frame our analysis and intend to contribute to. In section 3 we detail the methodology followed in the completion of this study. In section 4 we analyse the birth and early evolution of the new 'superconductivity' paradigm, hinting at the first economic implications for the computer industry. In section 5 we see the principle of presumptive anomaly at work: superconductors were technically superior to semiconductors, but the latter were far from unfolding their full potentiality. Section 6 illustrates the change of attitude towards superconductors while clarifying why, despite the demise of the superconducting computer project, the long run large-scale IBM investment was justifiable. Sections 7 and 8 provide further economic and non-economic reasons which justify such a massive investment. In the final section we discuss our findings and draw our conclusions[2].

---

[2] We wish to point out that when we wrote the first – conference – version of this paper in 2014, we could write that this was the first analysis *at all* of the Josephson computer case: we had gone through many books belonging to different vintages on the birth and evolution of computers (e.g. Golstine 1972, Flamm 1988, Ceruzzi 2003), and none referred to Josephson computing. However, while reworking on this paper we discovered that an historian of technology had been working, independently and in parallel, on the same topic (Mody, 2017, chapter 2); ours and his works are different and complement each other.

## 2. Theoretical background

The context of this study is that of science, technology and innovation, and the interrelation amongst these domains. In particular, central to the analysis are two fundamental concepts: *technological paradigm* and *presumptive anomaly*.

The concept of technological paradigm is widely known (Dosi, 1982). We also refer to two more contributions on the subject – Constant (1973) and Johnston (1972) – which complement Dosi's. The presumptive anomaly concept is due to Constant (1973, 1980) and applies to selected cases in which an incumbent technology is perfectly working, but it is expected to fail under "new or more stringent conditions"[3]; therefore, alternatives are sought even in the absence of any functional failure.

### 2.1 Technological paradigms

The technological paradigms with which we deal are those of the (incumbent) semiconductors technology and of (attempted alternative) superconductors technology. Following Dosi's definition:

"we shall define a 'technological paradigm' as 'model' and a 'pattern' of solution of *selected* technological problems, based on *selected* principles derived from natural sciences and on *selected* material technologies. […] We will define a *technological trajectory* as the pattern of 'normal' problem solving activity (i.e. of 'progress') on the ground of a technological paradigm." (Dosi, 1982: 152, original emphasis).

Once established, a paradigm becomes an accepted mode of technical operations as defined and accepted by the community of practitioners. Thus, following Constant (1973: 554) a technological paradigm is not just a device or a process, but is a coherent basis for action, which provides the ground for practice. It is a cognition, a particular shared way of perceiving a set of technology, which gets refined, improved and further articulated. In a similar vein, Johnston writes that a technological paradigm is: (i) an epistemological concept, in that it provides a guiding framework for the development of technology; (ii) a sociological concept, as membership of the community depends on the explicit adherence on the paradigmatic beliefs and principles; (iii) a psychological concept, for the technologist perceives the technological world through this framework, and organise his/her sense-data to conform to it (Johnston, 1972: 122).

The semiconductors technological paradigm established itself as dominant since the late 1950s as it synthesised the answers to the three aspects indicated in the quote from Dosi above. About *progress* along a trajectory, virtually everybody knows the 'Moore's law', i.e. the empirical law according to which the number of transistors squeezed in a given volume doubles every 18-24 months[4].

---

[3] These words are used by Constant (1980: 12) and, as he recalls, are taken from Kuhn (1970).
[4] See section 6 for more details on Moore's law.

However, the law is far from occurring automatically: in fact, cramming more and more transistors in the same volume requires better materials, better doping techniques, constantly renewed miniaturization technologies and many more improvements. The overlapping between the *scientific* paradigm of solid-state physics and the *technological* paradigm of semiconductors can be synthesised by the fact that at the very origin of modern electronics is the discovery of a new solid-state electronic component for which W. B. Shockley, J. Bardeen and W.H. Brattain were awarded the 1956 Nobel Prize for physics "for their researches on semiconductors and their discovery of the transistor effect". Shockley published quite a few papers on solid-state physics in the *Physical Review* and founded in 1956 the *Shockley Semiconductor Laboratory*. The emergence of a 'transistor techno-scientific community of practitioners' can be singled out by looking at the path which begins at the Bell Labs in the mid-1940s, and which led to Intel via the creation of the Shockley Semiconductor Laboratory and Fairchild Semiconductors[5].

When semiconductors were becoming dominant, the superconductor technological paradigm was in its early infancy, while the scientific paradigm – i.e. a theory capable of explaining superconductivity – was completely lacking. Superconductivity was detected at first in 1911, the first superconductive computer component, the cryotron, was devised in 1955, while a first *incomplete* theory concerning superconductivity was expounded in 1957. This is a case in which technology was ahead of science. This situation fits Johnston's words where he says that sometimes the emergence of a new paradigm in technology sets problems of understanding for scientists (Johnston, 1972: 128). In fact, a theoretical achievement was fundamental in opening the way to the realisation of much better superconductive devices (Josephson, 1962). In the mid-1960s the 'superconductor techno-scientific community' was emerging.

The search for an alternative paradigm implies dealing with a complex mix of dimensions which range from the relationship between science and technology to the question of how technological expectations form, from market considerations to non-market strategic government intervention – and this is not an exhaustive list.

A key question is thus the following: why was the superconducting paradigm actively pursued by IBM during the 1960s, when the semiconductors revolution was in full swing? The basic answer lies in the *presumptive anomaly* principle.

## 2.2 Presumptive anomaly

Paradigm shifts in technology may occur for different reasons. For instance, a technology generated in one field may be applied, with revolutionary consequences, to other fields: steam-engines were at first created to pump up water from coalmines, then they became the prime mover of the machinery of the Industrial Revolution in England, then they were applied to provide motion to ships so that steam-ships supplanted sail-ships. Steam technology was adopted by different sectors and thus adapted to many different uses.

---

[5] The history of transistors is not the remit of this work; a useful reference is Braun and Macdonald (1982).

The mechanism we are here interested in, however, is a specific one, that is the mechanism whereby a new paradigm is actively sought when the existing one perfectly satisfies the desired needs and ends. One such mechanism is that which has been named *presumptive anomaly* (Constant, 1973, 1980).

The original idea was developed with reference to the birth of the turbojet paradigm as opposed to the propeller-piston paradigm for aircraft propulsion. The successes of the 'old' propeller-piston technology were impressive in the period which ranges from the mid-1920 to the mid-1940: power of the engine increased ten times, from 350 hp to 3,500 hp, while speed doubled from 200 mph to 400 mph.

In the early days of aviation, aircraft performance and its improvement were based on empirical means. In the 1920s, however, aerodynamics became a scientific field characterised by mathematical rigour and communitywide acceptance. Soon after aerodynamicists started wondering what would happen to aircrafts at supersonic speeds, and concluded that the conditions of flight would change violently – and thus the propeller would not work 'normally' – as the velocity neared the speed of sound.

This was a rather awkward thing to do: supersonic speed means flying at about 770mph[6] – a speed that even today's civilian passenger turbojet aircrafts do *not* reach[7] –, and the top speed which could be attained when this conclusion was reached, was half that. Despite this, a solution to the 'problem' was sought, and brought to the turbojet revolution. In sum, the incumbent paradigm was experiencing important successful improvements, and the turbojet revolution was set against a background not of failure but rather against remarkable success. Thus:

 "Presumptive anomaly occurs in technology, not when the conventional system fails in any absolute or objective sense, but when assumptions derived from science indicate either that under some future conditions the conventional system will fail (or function badly) or that a radically different system will do a much better job. No functional failure exists; an anomaly is presumed to exist; hence presumptive anomaly." (Constant, 1980: 15)[8]

Parallels with the semiconductor and superconductor paradigms can be drawn: physicists warned that semiconductors were close to reaching a plateau already in the 1970s. Forty years later it is easy to say that those 'predictions' were wrong; whether IBM was successful or not requires a thorough analysis: at first sight, one would conclude that this is a failed innovation. However, we face a complex situation which does not allow a simple answer.

---

[6] The speed of sound is not a constant value as is affected by altitude and temperature; the figure above refers to sea level at 20 °C (or 68 °F) temperature.
[7] The *Concorde* and the Soviet-times *Tupolev TU 144* were exceptions and never became a widespread reality; the latter was in use only two years, while the former entered in service in 1976 and was dismissed in 2003.
[8] See also Constant (1973: 555).

It is interesting at this point to recall an 'opposite' mechanism with respect to the presumptive anomaly, namely the *sailing-ship effect* (De Liso and Filatrella, 2008). The latter is the phenomenon whereby the appearance of a new paradigm / technology (superconductors) stimulates improvements on the incumbent paradigm / technology (semiconductors). Thus, while presumptive anomaly leads to search for a new paradigm because the incumbent technology is expected to fail at some future point, the sailing-ship effect looks at paradigmatic competition the other way round: it is the emergence of an alternative paradigm which engenders intentional efforts to improve the incumbent technology.

A question might be: the spectacular advances observed in the semiconductor industry can have been engendered by the emergence of the superconductors paradigm?

This is not a redundant question. IBM was a major world player and when it announced worldwide – in specialised journals as well as in the widely-read magazine *Scientific American* – its Josephson computer project, it was sending a somehow threatening signal to the semiconductor industry: should IBM mass-produce in-house superconducting Josephson junctions, the world biggest computer producer would dispense with semiconductors.

*2.3 Science, technology and innovation: three interrelated domains*

The semiconductors and superconductors sectors are two big-science, big-technology, highly innovative sectors. The way in which the scientific and technological principles according to which these sectors emerged and evolved provides a very good example to appreciate the complex relationships existing amongst the three dimensions of science, technology and innovation.

Transistors and Josephson junctions are science-based devices, but the meaning of this phrase implies that science and technology are fully 'dependent' from one another.

Science, technology, and innovation are often portrayed as a sequence of interrelated stages constituting the engine of progress (Bush, 1945; Steinmueller, 1994). The literature on the topic developed rather organically since the late 1950s[9]. Amongst earlier studies, De Solla Price (1965) remarked a clear distinction between the domains of science and technology, positing the divergent nature of objectives and practices of science as creating and disseminating new knowledge against those of technology which consist in inventing new artefacts and processes with the objective to capitalise upon them on the marketplace. However interrelated, science, technology and innovation operate within distinct domains; each activity is driven by its own motives and follows its own paths (Brooks, 1994).

---

[9] The objective of this section is that of providing a theoretical background to our work and not an essay on the history of science, technology and innovation studies; we need to trace back important milestones in order to fulfil the general remit of this contribution. Investigations on science include publications by Polanyi (1956; 1962) and Popper (1959) amongst others studying the origin and the organisation of scientific endeavour, and Khun (1970) delving deeper into the dynamics of science. Whilst these publications are seminal in the study of science, similarly important works on science and technology began to emerge and were subsequently complemented by studies on innovation. To this day, this stream of research continues to constitute the foundation of STI studies.

Further contributions on the link between science and technology tend to present a dynamics where the boundaries between the two is blurring, especially in those emerging science-related technological sectors where artefacts and processes are substantially based on scientific discoveries and applications (Kranzberg, 1968; Freeman, 1982; Narin and Noma, 1985; Arthur, 2009; Rosenberg and Steimueller, 2013).

The relationships between the two domains are undoubtedly complex. Yet we can see how they originate from one common source: a 'problem' that inspires the search for a 'solution' which translates into scientific discovery in one domain and in new artefacts and processes in the other (De Solla Price, 1965; Agassi, 1966). However, we need to contextualise the scope of science and that of technology in relation to their dynamics (Arthur, 2009). This would release the implicit hypothesis according to which the dyad 'problem - solution' would be a singularity of both domains avulse from their contexts. This is not an isolated occurrence, and certainly has not limited reach; in science, it means the introduction of a new way of thinking, new modes of defining a problem and new models to characterise it, determining promising new avenues for further discoveries; Khun calls this a scientific paradigm (Khun, 1970).

Johnston (1972), Constant (1973) and Dosi (1982) draw explicitly from the Khunian approach a parallel to explain the dynamics of technological change. Set in a dynamic context, paradigms may evolve through cumulative accretions. This means that once a scientific or technological paradigm is established, a process of diffusion, refinement and institutionalisation of the state of the art develops across trajectories whereby variation and selection may engender individual dynamics[10].

By means of combining the more tangible aspects of a technology with its disembodied parts, a technological paradigm may also be underpinned by scientific knowledge especially in those emerging technological trajectories where the understanding of macro-phenomena is essential to technological development. Nonetheless, how potential innovations play out on the diffusion tail (i.e. in the market place) relies on mechanisms and dynamics that are not directly discernible from scientific advancement and technological change.

---

[10] See Anderson and Tushman, 1990; Vincenti, 1994; Antonelli, 1999; Mokyr, 2002; Mina *et al*., 2007; Arthur, 2009; Nightingale, 2014.

Nelson (2008) looks into the factors affecting the power of technological paradigms and argues that technological advances are based on the control of replicable practices. If science allows for such control, then the link between science and technology may be strong and technology may advance on the back of a robust scientific understanding: such is the case of scientific advances in fields such as mechanics, materials, biology etc., where the end game focuses on characteristics that may be embodied in physical artefacts. Nonetheless, controllable and replicable practices may evolve independently even though technological change is increasingly driven by the co-evolution of practice and understanding (Kranzberg, 1968; Freeman, 1982; Narin and Noma, 1985; Arthur, 2009).

The blending of the two domains has been studied by Foray and Gibbons (1996). The authors argue that there are particular patterns of knowledge generation that are tightly linked with technological breakthrough. When effort towards the advance of a technology (or group of technologies) relying on scientific knowledge in science-intensive domains operates in a context of knowledge availability, the problem-solution dyad is only a question of technical matter since the available scientific discoveries facilitate the construction of a framework to guide further research towards technological goals. On the other hand, whilst seeking to advance a technology without an underlying scientific knowledge base, advancement is sought in areas where cognitive domains are lacking and infrastructure, intermediate technologies and instrumentalities may not be relied upon for the technology to advance.

The case of superconductivity is challenging as at first it represents a case of technology ahead of science: the 'supercold' technological paradigm was being developed since the liquefaction of helium in 1908, and led to the creation of the first superconducting device, the cryotron, in 1955, without relying on a theory of superconductivity. A fundamental impulse to the technology, though, came after the *theoretical* physicist Brian Josephson published his paper on superconductive tunnelling (Josephson, 1962) which opened the way to the production of 'Josephson junctions' – which bear his name; yet, he never produced one.

A further fundamental challenge due to the study of superconductivity lies in the blurring of the distinction between science and technology: technology is more and more science-based, but science is more and more technology-dependent. This evolution in the relationship between science and technology is witnessed by the ubiquitous presence of physicists, in both the scientific and technological realm: they may be more theoretical or applied in kind, but their education and research activity necessarily synthesise scientific *and* technological paradigms.

## 3. Methodology

The objective of this work is to get theoretically sound insights from a case study which subsumes great scientific advancements, technological leaps forward and business-driven innovation spanning several decades and unfolding across diverse epistemological domains.

During the phase of planning of the study and scoping of the literature we found that our research objective could not be neatly analysed through the sequence consisting in basic scientific research followed by technological change and finally innovation (or business dynamics). These three domains, though playing an important role in our reconstruction, cannot be considered in progression from science through to technology and finally innovation as advocated at first by Bush (1945). Adopting this sequence would lead to an artificial and unsatisfactory reconstruction of the historical events thus misleading theoretical insights (Steinmueller, 1994).

We proceeded, therefore, by keeping the three areas of interest distinct without imposing any a priori relationships between them so that a historiographical narrative can be eventually reconstructed (Bruner, 1985; Polkinghorne, 1995). Ideally, a truly historiographical approach would consist in the reconstruction of the causal chain of events at least through causal integrability (White, 1965; Dray, 1971). However, it is well recognised that reconstructions of this sort would pertain to the ideal type, whilst practically unachievable. Following the Minkian tradition (Mink, 1966, 1968) we seek to reconstruct our story by considering and matching interrelated parts to delineate the network of complex relations emerging (Dray, 1971, Carr, 1986; Velleman, 2003).

Progressing along these lines we sought to identify the leading epistemological domains involved in the story, the key actors, their roles and relationships. In so doing, we could trace the milestones in each domain and their interdependencies including the idiosyncrasies which emerged. These are then put into context/perspective avoiding falling in the 'trap of hindsight' (Fischhoff and Beyth, 1975; Kamin and Rachlinski, 1995; Kahneman, 2011).

The primary sources of our reconstruction are traced back in the literature pertaining to superconductivity with applications to computers and constituent technologies. Following an established methodology on literature audit, snowballing, we proceeded by identifying the primary sources in the domains of interest (Greenhalgh and Peacock, 2005; Wohlin, 2014)[11]. The obvious primary keywords comprised "Josephson Computer/Junction/Effect/Tunnelling", "Supercomputer", "Superconductivity", and "Quantum Computer" as general search terms/phrases in standard databases such as the Web of Knowledge, Scopus and Google Scholar.

---

[11] Comparisons between snowballing literature research and systematic literature review have been carried out in several occasions, and in different domains. For examples in medical research see Greenhalgh and Peacok (2005) whilst in engineering see Jalali and Wohlin (2012).

Particular attention was given to physics journals, the IBM *Journal of Research and Development* and history of science and technology publications. After the research team identified primary sources – e.g. Bardeen et al (1957), Matisoo (1980), Novotny and Felt (1997), Likharev (2012), Gavroglu (2014) – we proceeded to a first, rough, reconstruction of the main domains of interest, identifying the phases of discovery in basic research, the technical and technological challenges, the business dynamics and the logic behind investment decisions. We sought to highlight the necessary antecedents, those necessary conditions for scientific discovery, technological experimentation and market dynamics using a retrospective outlook, which would link the various steps of our narrative. Finally, we looked at the interrelation between the three main domains, that of scientific discovery, technological development and business dynamics.

These phases were carried out accordingly to Greenhalgh and Peacock (2005) and Wohlin (2014); that is looking backward at the references contained in the primary sources in order to analyse the antecedents and reconstruct the basic narrative retrospectively. Subsequently, we searched forward to other sources citing or referring back to the primary sources in order to advance on our understanding of the process. This method was followed interactively on all primary and secondary sources in order to reconstruct the narrative of our story. Particular focus was placed on the idiosyncrasies emerging especially cross-domains where technological advances enabled new scientific discoveries or new discoveries would enable switching in technological applications or business opportunities. Early reconstructions of the story were circulated to selected scholars, presented at seminars, workshops and conferences held throughout 2013 and 2015 in various locations and with different audiences including economists and historians of information technology as well as research scientists, innovation scholars and physicists. Of course, the narrative was fine-tuned and revised in several occasions. Moreover, the authors held a workshop at the Department of Physics of the University of Salerno, Italy – where there exists a research group on superconductivity since the mid-1970s to which originally contributed the American physicist Robert Parmentier – in September 2015; here the authors benefitted from the comments and feedback by Arthur Davidson who participated in the IBM superconducting computer projects complementing our reconstruction with inside knowledge and insights (Davidson, 2015).

## 4. The superconducting supercomputer

The *superconducting*, or 'Josephson computer' revolution was publicly announced in March 1980 by IBM through its *Journal of Research and Development*. "Josephson computer technology: an IBM research project" is the explicit title of Wilhelm Anacker's article. This journal was known only to specialists. Another article, which granted worldwide acknowledgement of the new technology, was published two months later in *Scientific American* by Juri Matisoo (1980).

The revolution thus announced was based on *superconducting* materials as opposed to semiconductors. Behind the little change in the word, from *semi*-conductors to *super*-conductors, there is a paradigmatic change of technology – a new technology based on completely new principles and materials[12]. *Superconductors* needed the exploitation of new scientific and technological principles. This implied a radically new way to design and manufacture the basic components of computers.

## 4.1 The phenomenon of superconductivity: technology ahead of science

To understand the meaning of 'superconductivity' we have to begin with electrical resistance. All materials, including good electrical conductors such as copper, show some resistance to the passage of an electric current evidenced by energy loss. This general law does not hold for certain materials when refrigerated to temperatures close to the absolute zero (-273.15 °Celsius or 0 Kelvin). In these conditions, resistance drops to zero.

Getting close to absolute zero is, however, a difficult task. The technique relies on the liquefaction of gases, and in particular of helium. This process was the result of highly specialised laboratories using the most advanced technologies: there was no room for small scale experimenting nor for serendipity[13].

Kamerlingh Onnes liquefied helium in 1908 reaching a record-low temperature of -269° (4K), holding the monopoly on helium liquefaction until 1923. Once he had liquefied this rare gas, he started to investigate *systematically* the properties of materials at cryogenic temperatures. In 1911 Holst, a technician working in Kamerlingh Onnes laboratory, observed the sudden disappearance of mercury's resistivity at 4.2 K: superconductivity had been discovered (Gavroglu, 2014; Joas and Waysand, 2014; Matricon and Waysand, 2003). Before the phenomenon was detected, it was believed that superconductivity could not be reached because it needed the actual absolute zero temperature that cannot be reached for thermo-dynamical reasons. However, providing a theoretical framework to this discovery was not an easy task. Kamerlingh Onnes had proposed a formula in which resistance would gradually diminish together with lower and lower temperature, while competing theories predicted either a constant resistance or infinite resistance at the extrapolated 0K. Nonetheless, resistivity dropped suddenly to zero when the temperature reached the critical 4.2 K point (Mendelssohn, 1977), an option which had not been considered by any theory.

---

[12] Semiconductors constituted *the* paradigmatic component of the computer industry since semiconductors themselves defeated vacuum tubes in the 1950s and in 1980 the semiconductor-based technology was in full swing.

[13] The successful liquefaction of helium was the result of a technological competition between James Dewar, in London (UK), and Heike Kamerlingh Onnes, in Leiden, The Netherland (van Deft, 2014; Matricon and Waysand, 2003; Shachtman, 2000; Mendelssohn, 1977).

A first proper framework attempting a theoretical explanation of this phenomenon emerged only forty-six years after Holst's measurement. The path-breaking results were due to Bardeen, Cooper and Schrieffer who published an article entitled *theory of superconductivity* in 1957 (Bardeen et al. 1957)[14]. Interestingly, Bardeen was already awarded a Nobel Prize for Physics in 1956 for his "research on semiconductors and the discovery of the transistor" conducted at Bell Labs in 1946[15]. Bardeen was awarded a second Nobel Prize in 1972 and, this time, for his work on superconductivity.

Despite this success, today we still do not have a fully satisfactory theory of superconductivity (Schmalian, 2010). The lack of a fully consistent theory did not hinder the advances made on the technology of superconductivity. In the early phase of superconductivity research, both scientists and technologists believed that single chemical elements – such as mercury and a few more metals – had to be used and the main problem to solve was that of the purity of the elements. However, in 1986 Georg Bednorz and Alex Müller, two researchers working at the IBM Swiss research facilities, found that certain compounds, called perovskites, would become superconductor at much higher temperatures than pure metals, namely 35 K, or -238 °C – see Table 1 (Bednorz and Müller, 1987). Nowadays better superconductors, indicated as YBCO, have been devised; these become superconductor at 92 K or -181° C.

*4.2 Superconductivity research and the supercomputer*

IBM's interest on superconductors dates back to the mid-1950s, while heavy investments took place from the mid-1960s, which led to the 1980 announcement. The first applications of superconductivity were related to memory systems. This happened in a period in which there existed at least seven technological solutions for memory, yet no dominant design. In fact, in the early 1950s, memory systems could be based on magnetic tapes, magnetic drums, mercury delay lines, vacuum tube flip-flops, selectrons, William tubes and magnetic cores; a further possibility had emerged and was based on superconducting films (Gallagher *et al*., 2012). This early interest was not kept secret, and was actually made explicit in the fourth issue of the newly established *IBM*

---

[14] A theoretical work concerned with superconductivity had been published in 1950 by two Russian physicists, Vitali Ginzburg and Lev Landau; their work, however, did not explain the microscopic details of superconductivity. In 1959 another Russian physicist, Lev Gor'kov, demonstrated the complementarity between the Bardeen-Cooper-Schrieffer and the Ginzburg-Landau approach.

[15] In the 1956 Nobel lecture he points out that "the general aim of the program was to obtain as complete an understanding as possible of semiconductor phenomena, not in empirical terms, but on the basis of atomic theory" (Bardeen, 1956).

*Journal of Research and Development* in which one can find two articles (Crowe, 1957; Garwin, 1957) and a short communication (Kurtz, 1957) exploring *superconductivity*.

The early application of superconductors to computers' memory was based on the *cryotron*, a device that had been first proposed by D. A. Buck in 1955 exploiting the sharp transition from the normal to the superconducting state (Buck, 1956). As often occurs, early versions of new devices tend to be rather cumbersome and troublesome. Cryotrons were no exception, and were improved in many ways in the following years (Barone and Paternò, 1982; Brock, 2014). However, the technology of the time and some fundamental laws of physics did not allow a fast enough process of miniaturization of the device – a factor that would have been decisive.

IBM was working keenly on this early superconducting technology, devoting massive resources to it, supported by funding from the US Air Force: fast computing has always been of interest for the Department of Defence (NRC, 1999). Gallagher and co-authors speak of 100 staff employed, at various levels, on the project in the second half of the 1950s; however, the slow progresses led to a loss of interest, even though effort on superconducting continued (Gallagher *et al.*, 2012).

In the 1960s a few scientific and technological results were achieved. These granted superconductive electronics a new wave of interest. In 1962 Brian Josephson, a physics PhD student in Cambridge, England, published a theoretical key article on what was going to be later defined as the *Josephson effect* (see Table 2). As he explains in his Nobel lecture (Josephson, 1973) he wanted to provide an exhaustive general answer to a problem known to physicists as "tunnelling" – the passage of particles across a barrier whose energy is higher than the particles' energy. The treatment was particularly difficult in superconductors, in that electrons are connected in pairs. Josephson was able to predict that pairs can cross a non-superconductive barrier and a "supercurrent", in appropriate conditions, can flow between two superconductors – a *Josephson junction*.

While Josephson's was a theoretical result, the feasibility of a "Josephson junction" was demonstrated in 1963 by Anderson and Rowell – both working at Murray Hill Bell Telephone Laboratories.

Further encouraging results had been obtained within IBM where a Josephson junction was produced, which performed much faster than the original cryotron: "The combination of high speed, sharp threshold, and strong magnetic field dependence makes Josephson junctions attractive as logical elements" (Matisoo, 1966: 15).

| Table 1. Superconductivity – Key *technological* achievements | | | |
|---|---|---|---|
| Year(s) | Event / device | Who / where | Additional information |
| 1908 | Liquefaction of helium | Kamerlingh Onnes (Netherlands) | Competition with Londoner lab |
| 1911 | Superconductivity detected | Holst, at Kamerlingh Onnes Labs | The sharp transition to zero resistance was not predicted by any existent theory |
| 1955 | Cryotron | Buck, MIT Lincoln Labs (USA) | Use the normal-superconducting transition for binary logic element |
| 1957 | Further work on cryotron | Crowe, Garwin, IBM American Labs | Articles on the newly established IBM Journal of R&D |
| 1963 | Fabrication of a device to exploit tunnel of superconducting charge carriers (Josephson junction) | Bell Labs (USA) | Experimental confirmation of the existence of the Josephson effect |
| 1964-1983 | Josephson junctions circuit with photolithographic techniques | Leading group: IBM American Labs | Massive IBM investment on the Josephson technology |
| 1976 | Single flux quantum logic | Soviet Union (patented in USA) | Fast superconducting logic circuits – difficult to scale [see section 8 for further clarification] |
| 1987 | High-temperature superconductors | Bednorz, Müller, IBM Swiss Labs | Non-metallic materials; difficult to make reliable Josephson junctions |

| Table 2. Superconductivity – Key *scientific* achievements | | | |
|---|---|---|---|
| Year | Event | Who / where | Additional information |
| 1950 | Phenomenological description of superconductivity | Ginzburg, Landau (Soviet Union) | Does *not* explain the microscopic origin of superconductivity |
| 1957 | First theory of superconductivity "BCS" | Bardeen (USA), Cooper (USA), Schrieffer (USA / UK) | Bardeen worked at Bell Labs in the years 1945-1951 |
| 1959 | Proves that the Ginzburg-Landau equations are a consequence of Bardeen-Cooper-Schrieffer theories | Gor'kov (Soviet Union) | Microscopic derivation of the Ginzburg-Landau equations |
| 1962 | Theoretical explanation of superconducting tunnelling | Josephson (Cambridge, UK) | Based on the BCS theory |

An important consequence of the latest IBM achievement was that superconductors, when combined to form a Josephson junction, could be used not only for memory systems but also for *logic circuits*, the computer's central processing unit. Thus, now the two key components of the computer's architecture, i.e. memory and logic circuits, could be based on the same scientific principle and technological devices, i.e. superconductivity and Josephson junctions, respectively.

These results prompted renewed interest, meaning heavy R&D expenditure in superconducting computers. Despite semiconductors progresses, at the beginning of the 1980s it looked like superconductors could eventually win the market: "Josephson tunnelling devices … could indeed be switched very fast and could be competitive with projected semiconductor integrated circuits" (Anacker, 1980: 108). In 1980 the world had to know that a new technology, based on a new paradigm, was getting centre stage in the computer industry.

## 5. Presumptive anomaly at work: superconductors in a world of semiconductors

One question has been voluntarily omitted up to now: why was there such an interest in this 'strange' technology which implied the creation of a new paradigm? Was there anything wrong in the semiconductor technology, or was it evolving so slowly that it could no longer meet increasing calculation needs? The answer to the latter question is "no".

Yet the struggle continued. In order to build a 'Josephson computer' new principles and radically new technologies needed to be developed. Many technical difficulties had to be overcome, and many sources of cost considered.

By 1980 the semiconductors industry had long identified silicon as the "right" material, while in superconductors such a material had just been identified: not all metals become superconductors at near-absolute-zero temperatures, and the selection of niobium instead of lead required a lot of experimenting.

A fundamentally obvious aspect consists in keeping the temperature close to absolute zero. The use of liquid helium to maintain the ultra-low temperature, was – and still is – very expensive. It adds very heavily to the costs not only to test and mass-produce the devices, but also to simply keep the computer running.

The materials must withstand temperatures which range from -270 °C to +40 °C or more, that is more than 300 degrees thermal cycle. Given that the Josephson junction is made of two superconductors separated by a non-conducting barrier, the latter material can be characterised by a different thermal expansion/contraction. Working on a nano-scale this differential expansion or contraction can jeopardize the functioning of the junctions (Mukhanov, 2014).

However, the new technology had quite a few strengths, while it could also benefit from some of the techniques used for producing semiconductors devices (Anacker, 1980: 109 and Matisoo, 1980: 63). The realisation of Josephson chips and circuits could partly be based on the same lithographic resolution of semiconductors.

There were a number of reasons to believe that superconducting electronics was superior to semiconducting devices. Josephson junctions were considered promising elements for 'artificial thinking' – or 'neural networks', as they were called in the 1960s (Cull, 2006).

It was by no means obvious that neural networks could work on semiconducting devices, so it looked more promising to build special-purpose devices, named *neuristors*, which would be the man-made technological version of neurons, based on Josephson junctions[16]. The program to build such special-purpose devices continued for all the 1960s (Parmentier, 1969) and research at the Wisconsin University was popular enough to be the basis of a science fiction book (Zelazny, 1976).

It is interesting to notice that in 1967 power dissipation was clearly seen as a limit for semiconductors, and it still is an important concern nowadays. It is perhaps even more interesting to notice that the economic analysis was based on the expectation that the cost of semiconductors would become *higher* than the cost of superconductors.

In fact, semiconducting devices are made of 'doped'[17] silicon (or germanium). In contrast, Josephson junctions were at first made of inexpensive lead. Nobody in the 1960s could predict that a transistor (three differently doped pieces of silicon glued together) could cost 0.004 US$ by the mid-1970s. It was reasonable to believe that a few billions one-millimetre Josephson junctions, made of lead or aluminium, could be much cheaper. Therefore, the expectation devoid of the hindsight effect, was that cheap superconducting components would more than offset refrigeration costs.

From the early 1960s semiconducting transistors were the dominant technology, and technical progress took place at an accelerated pace. The market for transistors was expanding dramatically also because transistors found applications in the telephone industry (Ceruzzi, 2003). Thus, giant companies producing transistors could expand their R&D activities through increasing resources coming from exponentially growing market sales. Despite this, R&D on superconducting devices was kept alive because in principle it was superior in terms of performance and could partly benefit from progresses made in the semiconductor field, such as the photolithographic and epitaxial techniques allowing greater miniaturization and concentration of junctions in the same volume. One main difference between the *super* and *semi*conductors was thus in terms of resources available to carry out R&D: semiconductors had a strong advantage.

---

[16] The term neuristor was introduced by Crane (1962). It describes distributed active electronic devices mimicking nerve axons in the propagation of electrical pulse. In his paper, the author discusses in detail a technique of stability analysis applicable to neuristor waveforms through a Josephson junction device which offers some promises for truly practical physical realization (see also Parmentier, 1967).

[17] *Doping* consists of the insertion of small and controlled quantities of material in an otherwise pure semiconductor; it is a delicate process that few foundries are, still nowadays, capable of performing.

At the very beginning of the 1980s, photolithography was available for semiconducting and superconducting devices, with the same spatial resolution. Together with other significant improvements of superconducting technology (such as the tri-layer all Niobium fabrication method), it was realistic to think that superconductors and semiconductors were competing upon equal terms.

Superconducting computers, though, were expected to have specific technological and economic advantages. IBM, through the works of Anacker (1980), Matisoo (1980) and other publications, proposed the scenario that the expected performance of the superconducting computer would be at least 10 times faster than that based on semiconductors – a 100-fold higher computing rate could be conceived – and the higher costs to keep supercomputers temperature close to absolute zero would, in time, be overcome.

Reasoning in terms of comparative trends characterised by continuous increase in performance, squeezing more and more transistors in the same volume would necessarily lead to a stage in which overheating would constitute a problem for the incumbent technology.

The estimates proposed by Anacker (1980) with regard to a mainframe capable of performing 250 million instructions per second, the best semiconductor technology of the time, was that it would dissipate about 20 KW of power: an amount impossible to extract from highly concentrated circuits. In other words, even considering a tenfold gain in efficiency of heat dissipation, the temperature would be so high as to compromise the working of the standard semiconductor-based machine. Therefore, expensive cooling technology would be needed to keep the mainframe running. On the other hand, the proposed ultrahigh-performance of a superconducting mainframe was estimated to dissipate less than 10 W. "It appears, therefore, that Josephson technology may, at present, be the only one with potential for construction of mainframes of the ultrahigh performance" (Anacker, 1980: 109).

This means that refrigeration costs would not be crucial for they would emerge in any case also for the semiconductor-based mainframes. Furthermore, the new technology would be systemic in that the materials and the overall architecture would be based on these new principles operating at low temperature making "overhead for refrigeration" not a disadvantage for large- and medium-scale mainframes.

Two more advantages of superconducting computers would be reliability and longevity of components. In fact, the materials used are selected from the very beginning to cope with supercold temperatures, so they would not have to adapt ex post to refrigeration – a problem specific of semiconductors. Longevity would be guaranteed by the low temperature.

Given the intrinsic superiority of superconductors, fewer circuits might be required, thus reducing the manufacture of chips and testing costs (Matisoo, 1980: 63).

A further technical aspect which adds to the superiority of Josephson circuits consists of the fact that these devices are indifferent to the direction in which the current flows through them, while the same is not true for semiconductors (Matisoo, 1980: 56-57).

Last but not least, all the contributors to the development of the superconducting computer pointed out that R&D-cum-learning-by-doing would give rise to forms of progress similar to those experienced in the semiconductor industry.

Summing up: since the 1960s semiconductor-based computers were developing at an impressive rate, and yet the expectations in the late 1970s were that a functional failure would 'soon' be reached – another way to say that there existed a presumptive anomaly. In hindsight we know that 'soon' was not so close in time – but we are all capable ex post to say that the expectations were wrong.

## 6. From massive investment to the demise of the superconducting supercomputer

"On 23 September [1983], IBM ended its attempt to build a high-speed, general-purpose computer whose guts would be logic and memory chips made of superconducting Josephson junction switches": thus wrote Arthur L. Robinson in the November 1983 issue of *Science*.

Within three years we have a complete change of attitude on superconductivity in the computer industry. In hindsight it may be easy to classify the investments made by IBM in the new technology as a waste of money. However, the company's bet on superconducting computers was justifiable on both technological and economic ground.

Any investment is based on the expectations formed at the time when the commitment is made. Expectations concern the inextricable mix of technological and economic future conditions as well as strategic considerations: how do we prioritize emerging technologies (Mokyr, 1998; Kim *et al*., 2010)?

The semiconductor-based computer industry was undoubtedly one of the most progressive ones at least since the 1960s, so that IBM had to shoot at three fast-moving targets, that is: (i) the evolution of the performances, (ii) the costs structure, and (iii) the size of the market for computers.

R&D and mass-production would determine performance and cost structure, and thus market-size, while the latter would feed back into the former dimensions; this dynamics for IBM was extremely important. In fact, from the early 1960s IBM had been a dominant player in this industry. Such was its importance that it was accused of abusing its dominant position in this market, and from 1969 to 1982 was subject to one of the most important antitrust cases ever. The case was dropped by the US Government because the company demonstrated that its near-monopoly had been obtained and maintained through "superior skill, foresight and industry" (Fisher *et al*., 1983). Nonetheless, this dominant position implied fat profits: IBM was *the* most profitable American company in 1983 and had always been second or third since 1970 according to the *Fortune 500* ranking. Despite this dominant position, based on a well-established technology – semiconductors –, the company decided to develop also an alternative promising technology – superconductors.

Was this an absurd strategy? Let us consider as a starting point the so-called 'Moore's law' formulated by Gordon Moore in an article published in 1965, whose title "cramming more components onto integrated circuits" is quite telling (Moore, 1965). He proposes an estimate of the reduction of manufacturing transistors costs together with the exponential growth of the number of components per circuit (a doubling of the number of transistors per circuit every *12* months). For the decade 1966-1975 this prediction showed to be accurate. When IBM announced its Josephson computer programme Moore's law had been at work for a few years already and it had slowed down to the version which sees a doubling of transistors per chip every 18 or 24 months. The transistors technology had evolved into integrated circuits (1959) and then into microprocessors (1971).

There was no estimate of the upper performance limit of semiconductors, though all agreed that a limit does indeed exist, and it is fixed by the laws of physics. What has happened is that this limit has been shifting upwards through time, as technology showed that the supposed limits imposed by physics had to be revised. Provocative research has shown that the upper bound could be placed almost as high as one wishes (Krauss and Starkman, 2004). Yet, in the late 2010s we might be close to the final limit because we are working on such a nanoscale – single atom – that a phenomenon known as superparamagnetic effect might appear in the magnetic memory: the memory chip would not be able to keep stably the 0s or the 1s on such a small scale (Economist, 2016). At the *present stage* of scientific and technological knowledge, that limit is insurmountable. Perhaps this is the ultimate limit, but other 'ultimate limits' were envisaged before, and have all been overcome[18].

Was all of this predictable? Moore's law was bound to saturate, and timing was central. Ex post it is easy to say that IBM miscalculated the time scale of the duration of the validity of Moore's law and that it made its bet on the new technology thirty-five years too soon. However, when IBM was actually working on superconductors the hypothesis of the "law" no longer holding for many years was not unreasonable.

---

[18] For example, it was believed that photolithography was ultimately limited by the wavelength of light, but the advent of electron microscopy demolished the limit through using electrons whose wavelength is much shorter. Nevertheless, some limit or unbeatable difficulty will occur.

In the overall emergence of expectations, one must also consider the fact that steady improvements could be expected to occur in mass-production of Josephson junctions since their production shared the same lithographic technology as semiconductors and some specific design characteristics might be "easier in some ways than the manufacture of semiconductor components" (Matisoo, 1980, p. 63). In other words, it was perfectly rational to believe that superconductors would improve at the same rate as semiconductors, if not faster. Thus, the relative distance in terms of performance between the two technologies would stay the same over time, highlighting a significant advantage for superconductors. Furthermore, sooner or later the heat problem would have emerged on the semiconductor side, thus further shifting the balance in favour of superconductors.

One more reason to invest in superconductors lies in the possibility of a type of cost reduction that would have been superconductor-specific. In fact, so-called high-temperature superconductors were actively sought. These would use cheap liquid nitrogen, as opposed to expensive liquid helium[19]. The use of nitrogen would have substantially reduced the cooling costs[20].

Despite the fact that IBM officially abandoned the Josephson computer project in 1983, its laboratories were still devoting some efforts to superconductivity. In particular,

"[Alex Müller] had worked on strontium titanate and related perovskite compounds. … But not until a twenty-month sabbatical spent at the IBM Yorktown Heights Laboratory did Müller become familiar with the intricacy of superconducting research. He was particularly fascinated by the IBM Josephson computer project, a \$ 300 million effort to build a fast computer based on superconducting switches. The computer project was later abandoned, but knowledge that several complex oxides were superconducting led Müller to believe it might be possible to find new compounds with unprecedented critical temperatures" (Nowotny and Felt, 1997: 30-31).

## 7. Expectations on the evolution of the market for computers

One crucial question emerges here: which market are we talking about when we talk about computers in the early 1980s? IBM had been working on superconductivity since the mid-1960s, and when it formally announced the Josephson technology project in 1980, *mainframe* computers were still at centre stage. By then the personal computer revolution had already started, but the fact that computers would become as widely diffused as washing machines was not to be taken for granted. Once more expectations play a key role.

---

[19] The importance of active research in high-temperature superconductors should not be underestimated: it is not a simple coincidence that in 1987 two IBM researchers, Georg Bednorz and Alex Müller, were awarded the Nobel Prize for physics for having found materials at comparatively 'high temperatures'. The passage from 4 K to 23 K took 75 years, while the passage from 23 K to 92 K happened suddenly between April 1986 and February 1987 (Müller and Bednorz, 1987).

[20] There exists a nonlinear relationship between cooling a material and the energy necessary to cool it. Getting closer to absolute zero by one degree at a time would add to the costs at an accelerating rate: to decrease the temperature from 80 K to 79 K (the range of temperature of high-temperature superconductors) costs 10 times *less* than to decrease the temperature from 5K to 4K (the range of temperature of old traditional low-temperature superconductors).

The market, was expected, will be segmented into two asymmetric halves.

The rich half: that of mainframes for public administrations (from central ministries and departments to local councils), universities, banks, insurance companies, airlines, publishing houses, newspapers, hospitals and any business which needs handling of large databases, which as a minimum means the accounting division of any firm. The list above is already quite long and, one notices, comprises businesses or activities belonging to the service sector. The list has to be enlarged to take into account many manufacturing processes needing large computing facilities to run the overall production process such as flexible manufacturing systems in which numerically controlled machines perform the whole manufacturing process. Numerical control, i.e. mainframe computerised work processes, already was a dominant technology in the 1970s (Reintjes, 1991).

If we can expect a rate of progress in superconductors as in semiconductors, mainframes will become more and more affordable so that smaller and smaller firms may afford one.

The second half of the market is a poorer one. Made of small users and giving rise to little profits.

Was it such a bizarre thing to hold this expectation in the 1970s? Our answer is simply no. In hindsight, we know how the situation evolved after 1980, with the explosion of the semiconductor-based personal computer market. A further (and unexpected) change occurred in the 1990s with the advent and diffusion of the internet. Personal computers, laptops and hand-held devices have spread in offices, gradually into homes and now are personal items. In the 1960s and 1970s, many people were just laughing at the idea that everybody could use a personal computer (Platt and Langford, 1984).

However, this is only part of the story of the semiconductors permanent revolution and continuous improvements. In fact, through time the semiconductor microchip has emerged as *a general purpose technology* (a phrase made popular by Bresnahan and Trajtenberg, 1995). Semiconductors have benefited from continuous growth of the market, thus granting nearly limitless resources to semiconductor-producing firms allowing them to work on a scale whose dimension is difficult to estimate.

Convergence between various ICTs sectors – just think of the entertainment industry, from television to cinema, music, videogames, smartphones etc. – has amplified the need for semiconductors, and has given rise to knowledge, technological and application convergence which coevolve and generate innovations (Hacklin *et al.*, 2009). Furthermore, microchips can be found everywhere, from cars, to fridges, washing machines, lifts, stereo sets, medical apparatus, heating systems, air conditioning. Basically, wherever there is electricity one finds microchips.

## 8. Beyond 'simple' market and economic considerations

In the previous sections we have emphasised technological and market considerations. However, the history of computing is driven by other factors with repercussions on the whole industry. These include military hubris, the ostentation of technological superiority and the willingness to become part of the technologically advanced societies. In other words, non-economic factors have heavily affected this sector since its very beginning.

It is difficult to quantify the involvement of the American government and its many Agencies – from the Department of Defense to NASA – in direct and indirect funding of R&D aimed at improving computing even for insiders (Zimet *et al*., 2009). The lack of publicly available micro-data, together with problems of consistency in data available, makes it impossible to disentangle the financing of specific projects.

However, some studies provide analyses and sectoral data which make clear the fundamental importance of the American Federal support to computer science research. Funding has concerned systematically basic research, human resources and research equipment, in both computer science – as 'strictly' defined: theory of computing, design, data storage and manipulation, systems analysis, programming languages – and electrical engineering – which includes communication technologies, semiconductors, superconductors, electronic circuits as well as in electric power (NRC, 1999). Through time observers have systematically emphasised that the American information technology industry thrives to a large extent on the basis of an extraordinary track record of public research funded by the Federal government since the 1940s (Goldstine, 1972; Flamm, 1987; NRC, 1995; Zimet *et al*., 2009, Mazzucato, 2015).

IBM has systematically co-operated with many of the American governmental agencies interested in faster and faster computing, and superconductors were part of the game. By the second half of the 1970s, when the IBM Josephson project was running, roughly one-third of total computer-related research funding came from Federal funding (NRC, 1999: 60). One particular datum is also of interest: in 1963 the government funded 35% of IBM's R&D in computing (NRC, 1999: 59). Supercomputers – whatever technology they are based on – are important for many reasons, from driving missiles to predicting a hurricane's path and optimization of encryption/decryption operations. Given their strategic importance, economic considerations may have been sometimes set aside.

Another player which did invest heavily in superconducting supercomputers was Japan. Fundamental was the Ministry of International Trade and Industry, or MITI – METI since 2001 – (Takagi *et al*., 2008; Fujimaki, 2012). The world realised the importance of MITI in the 1970s and 1980s, when it was very effective in promoting the Japanese hi-tech sectors by means of various tools, a constant one being carrying out and subsidising R&D on a large-scale. Between 1975 and 1984 MITI created and sponsored four different consortia aimed at catching up and possibly overtaking the United States in computer technologies (Callon, 1995). One of the consortia aimed explicitly at building the world's fastest computer, and considered Josephson junctions as a means to achieve this result. This consortium, which lasted until 1990, comprised MITI's Electrotechnical Laboratory (known as ETL) – which allocated nearly US$135 million to the project – Fujitsu, Hitachi and NEC with matching funds.

The results achieved were defined "substantial, sustained and highly productive" in a report written by an American panel which was visiting Japan to investigate progresses in electronic applications of superconductivity (Rowell *et al.*, 1998). In fact, the Japanese had fabricated a working superconducting all-niobium technology microprocessor which contained 24,000 Josephson junctions[21].

Japan's efforts also involved the university system in order to develop basic knowledge of superconductivity as well as to search for materials which would become superconductors at relatively high temperatures. By the late 1980s two different roads were thus open: low-temperature and high-temperature superconductivity (Merali, 2012).

Interest in superconductivity went thus on without interruptions: Japan, in fact, heavily financed two more consortia. The first was the Superconducting Sensor Laboratory (SSL) (1990 to 1996); the second started immediately afterwards, in 1997, the International Superconductivity Technology Centre (ISTeC). The latter – which despite its name was fully Japanese – was at the time the largest consortium devoted to the development of superconductivity in the world (Rowell et al. 1998). Further efforts are being carried out by METI since 2006 in order to produce and apply *superconducting* quantum interference devices, or SQUIDs, and single flux quantum devices, or SFQ (Tsukamoto, 2008).

Also the Soviet Union – uninterested in market aspects by definition – was interested in the technology. As Likharev (2012) recounts, in the 1960s the Soviet government established a large research facility (NIIFP), in order to perform reverse engineering of foreign semiconductors and study other technologies, including superconductors. Likharev – himself an important contributor to the field of superconductivity – was working at the Moscow State University and devoted much effort to fabricate superconductor "bridges" at first with a small group of researchers. The group had some early success and was granted two patents[22] for their original work in devising thin-film

---

[21] For a complete technical explanation see Kotani *et al.* (1990).
[22] US Patents #4,164,030 and #4,178,602.

Josephson cryotrons. In the early 1970s the Moscow State University team and NIIFP started to co-operate. A few years later further co-operation was established with another research team at the Institute of Radio Engineering and Electronics of the Soviet Academy of Sciences. In 1986 these groups created the world fastest circuit, operating at up to 30GHz, and predicted the possibility to increase the clock frequency to a few hundred GHz (Likharev, 2012).

As one can see, the efforts had been financed for over twenty years, and further resources had been granted just before the breaking up of the Soviet Union.

Summing up, in the computer sector, we have a situation which has occurred rather often in technologies held strategic, that is the fundamental role of governments in favouring R&D and early development and – in market economies – the subsequent further development promoted by firms.

The role of companies competing in open markets has been important in the development of semiconductors. In fact, this technology moved from the infant to the take-off stage, new Schumpeterian firms entered the market and, by 1957, 64% of the total semiconductor market had been captured by these new firms (Braun, 1981: 74).

Superconductors did not reach the critical mass in the early 1980s as IBM had expected and, despite the improvements obtained by IBM, the Japanese and Russian players, interest in superconducting computers lost momentum.

Nonetheless, this is not the end of the story of superconducting for three reasons. First, the Japanese took over and produced a working superconducting microchip in 1989, while some successful work had taken place in the last years of the Soviet Union. Second, the IBM Josephson computer project may actually have *never* been completely abandoned, but just reduced in scale[23]. Third, despite the fact that *binary* superconducting supercomputers were never produced beyond the superconducting microchip prototypes, thirty years after the closing of the IBM project, superconductivity emerges as a key component of the *quantum* computers of D-Wave (Stajic, 2013). Whether quantum or not (Shin *et al*., 2014) there remains the fact that this is an ultrafast computer whose speed depends on cryogenic temperatures and the Josephson's effect.

Moreover, further technological achievements based on superconductor technologies have been attained during the last thirty years: magnetic resonance imaging, magnetic levitating trains, highly sensitive measurement instrumentation and other devices would not exist without superconductors (DeBresson, 1995; Malozemoff *et al*., 2012).

---

[23] Dorojevets and colleagues write explicitly that superconductor technology and processor design continued behind the scenes. They refer to a hybrid technology reaching a processing capability of one petaflop, or $10^{15}$ operations per second. This speed was deemed necessary "in the national interest" - the authors explicitly acknowledge "support by the US Department of Defense" (Dorojevets *et al*., 2004; Brock, 2016).

Furthermore, scientific achievements have emerged. Through the study of supercold materials the Bose-Einstein *condensate* – a state of matter predicted in the 1920s by the two scientists which was thought to be impossible to observe – could finally be produced for the first time.

## 9. Discussion and conclusions

To conclude we consider three major points. The first concerns the question whether superconductors can be considered as a failed innovation. The second concerns the way in which presumptive anomaly works. The third concerns the relationship between science and technology.

Presumptive anomaly is the theoretical tool which helps us to explain the creation and development of a new paradigm, based on superconductors, aimed at substituting the incumbent paradigm, based on semiconductors, in the computer industry. The fundamental basic mechanism indicated by Constant is clearly at work: the incumbent technology is perfectly working, but expectations on some future failure of the same technology stimulate the search for radical alternatives.

The question which immediately emerges is whether that of superconductors can be classified as a failed innovation.

The scale of investment by IBM, corroborated by support of US Federal government, was impressive: a research group of 100 to 125 persons worked on the Josephson computer project between the mid-1960s and the early 1980s. R&D expenditure by IBM alone exceeded US$ 20m per year during that period (Logue, 1998; Davidson, 2015; Mody, 2017). The efforts devoted to Josephson computers were thus enormous.

However, despite IBM 1980 public announcements and research efforts, doubts were present; there was competition *within* IBM between the Josephson and the semiconductor departments; furthermore, there existed competition within the Josephson IBM community between the American and the Swiss laboratories (Davidson, 2015).

In 1981 a new Josephson project leader, J.C. Logue, was appointed *en lieu* of Anacker. The executive technical unit lead by Logue was made up of 15 engineers and scientists, only three of whom recommended that the program be continued. IBM decided to give the Josephson technology two more years before taking a 'final' decision, which came in 1983. As Logue puts it: "the 125 members of the Josephson program could not compete with the thousands of people around the world which were engaged in improving the silicon technology" (Logue, 1998: 66-67).

The last quote is crucial: the superconductors-based paradigm did not reach the point of revolution, which comes when a significant minority of the relevant community switches to the new paradigm, thus making its development self-supporting (Constant, 1973: 556). However big, IBM alone could not compete with the semiconductors' community, in which there also existed giant companies such as Intel – founded in 1968 – committed to keeping Moore's law going for decades by heavily investing in R&D.

Thus, given that Josephson computers never became a widespread reality, should we conclude that this is a failed innovation? Not necessarily. The right classification might be that of a "false-failed innovation" which is defined as "a technology that is examined and discarded but that gets a second chance under other conditions and succeeds" (Wilmoth, 2000: 51).

In fact, the new generation of (quantum?) computers launched in the 2010s makes use of Josephson junctions, and more agencies (ODNI, 2014) and high-tech companies (Hypres, 2015) are back, re-investing in superconductors applied to computers. Then, should the new paradigm reach the critical mass, it will be interesting to see if the 'sailing-ship effect' will take place. The latter effect, however, was not present during the years when Josephson junctions were experimented; we cannot say that superconductors were displaced by the improvements of semiconductors *caused* by the emergence of superconductors themselves.

Incidentally, in the overall evaluation of the superconductors' paradigm until today, one should not forget the other applications of superconductors, and in particular magnetic resonance imaging, or MRI: these machines definitely constitute a widespread reality, as they are present in every hospital and in many other medical facilities.

The second point we want to raise, concerns the working of the presumptive anomaly principle. In the original idea by Constant, the principle depends crucially on *scientific* insights which indicate that under some future conditions the incumbent system will fail or function badly. In the case of semiconductors, the indications need not come from scientists: it was clear also to technologists that a problem of heat would emerge by squeezing more and more transistors in a given volume as it was clear that miniaturization to a nano-scale would create serious manufacturing problems. The step from the identification of these two basic problems – which will 'soon' invalidate Moore's law, and will become insurmountable at a later stage – to the search for a new radical alternative is somehow 'natural'. Thus, presumptive anomaly does not need to have its roots exclusively in science.

Furthermore, in our story, the drivers of change can be only partially explained by scientific, technological and business ambitions. Though, forming the underlying reasons to the sequence of events that we reconstructed, these cannot be unequivocally subdivided and classified. Factors external to these domains can be brought about to justify the enormous commitment to superconducting development. In particular, the interest of governments and their active intervention in computer technology is a constant feature of the technology itself. The very birth of the computers field was characterised by heavy public investments: the history and realisation of the first monster computer built in the 1940s, ENIAC, under the very generous American government aegis, is widely known (Goldstine, 1993). The drivers to embark in a difficult process cumulates momentum feeding ambitions from various fronts, including ideological/national hubris, international competition and technological foresight. Operating within this environment, the actors of our story acted under the assumption that market (or national) superiority would provide strong incentives in pursuing a project that has not brought marketable results in over 50 years, but it may eventually develop further.

The third point we focus on, concerns the relationships between science and technology. This relationship changes through time, and varies from field to field. It is not possible to identify an invariant universal pattern. Also, the relationship is sometimes 'intermediated' by governments, in the sense that the latter can steer the overall system, through massive investments and involvement of private companies, towards specific technologies and scientific areas.

The field of superconductivity is *big technology* and *big science* from the very beginning. Kamerlingh Onnes cryogenic laboratory, in which superconductivity was first observed, was the realization of a technological project of unprecedented scale. Liquefaction of helium in 1908, which made possible the observation of superconductivity in 1911, was a major *technological* success. The first cryogenic devices aimed at building computer memory circuits were built in the mid-1950s without much theoretical guidance: we have a case in which technology is ahead of science – and we have to note that this story disproves the idea that it was only in a distant past that through ingenuity and dexterity one could 'make a steam engine without knowing the principles of thermodynamics'.

What we have here is a situation in which science chases technology, trying to provide the explanation of the principles underpinning certain phenomena. Technology and science are still sometimes satisfied by answering different questions. Technology asks: Does it work? Is it reliable? Can we make it smaller and faster? Science asks: *Why* does it work? What does explain the behaviour of the phenomenon under investigation? Which is the predictive power of my theory? The understanding of scientific principles, though, is sometimes a necessary condition for technological improvement: it was only after Josephson's 1962 theoretical paper that predicted the properties of superconducting junctions, that through new experimental research junctions themselves could be fabricated in ways that could match and overtake, in terms of technical performance, semiconductor devices.

To conclude, we wish to make two remarks, one general and one specific. The general one concerns the fact that our work points once more to the complex, nonlinear, characteristics of the innovative process: scientific endeavour, technological advancements, government's 'interference' and business dynamics intertwine. The specific remark concerns the development of superconductivity: many technological and scientific results have been achieved. Superconducting phenomena and apparatus have been systematically exploited and used in MRI, special antennas, energy storage, highly sensitive instruments, particle accelerators and magnetic levitation – besides in the Josephson junctions. All of these are the results of intentional efforts. Superconducting computers, though, have not become a widespread reality. However, the technology is there, and semiconductors are bound to saturate their improvability: maybe, a new chapter on the paradigm of superconductivity will have to be written in the near future.

**Acknowledgements**

A preliminary version of this work was presented at the European Forum for Studies of Policies for Research and Innovation (EUSPRI) conference on "Science and Innovation Policy: Dynamics, Challenges, Responsibility and Practice", Manchester, UK, 18-20 June 2014. Comments by the participants are gratefully acknowledged. Conversations with Jonathan Aylen, Hugh Cameron and Giovanni Costabile helped to improve the manuscript.

A special acknowledgement is deserved by Arthur Davidson who participated in the IBM superconducting computer project and with whom we had a long encounter in September 2015.

Last but not least, we wish to thank the two anonymous Referees for their suggestions, which undoubtedly helped to improve the content of this work. None of the people above bear responsibility for any remaining error.

**References**

Agassi, J. (1966), 'The confusion between science and technology in the standard philosophies of science', *Technology and Culture*, **7**(3), 348-366.

Anacker, W. (1980), 'Josephson Computer Technology: An IBM Research Project', *IBM Journal of Research and Development*, **24**, 107-112.

Anderson, P. and M.L. Tushman (1990), 'Technological discontinuities and dominant designs: A cyclical model of technological change', **35**(4), *Administrative Science Quarterly*, 604-633.

Anderson, P.W. and J.M Rowell (1963), 'Probable Observation of the Josephson Superconducting Tunnelling Effect', *Physical Review Letters*, **10**, 230-232.

Antonelli, C. (1999), 'The evolution of the industrial organisation of the production of knowledge', *Cambridge Journal of Economics*, **23**(2), 243-260.

Arthur, W.B. (2009), *The Nature of Technology. What it is and how it Evolves*. Free Press: New York.

Bardeen, J. (1956), 'Nobel Lectures', Physics 1942-1962, reprint 1964 Elsevier Publishing Company: Amsterdam.

Bardeen, J., L.N. Cooper and J.R. Schrieffer (1957), 'Theory of Superconductivity', *Physical Review*, **108**, 1175-1204.

Barone, A. and Paternò, G. (1982), *Physics and Applications of the Josephson Effect*, Wiley: New York.

Braun, E. (1981), 'From Transistor to Microprocessor', in T. Forester (ed), *The Microelectronics Revolution*, MIT Press: Cambridge, Mass., 72-82.

Braun, E. and S. Macdonald (1982), *Revolution in Miniature. The History and Impact of Semiconductor Electronics*, second edition. Cambridge University Press: Cambridge.

Braun, H-J. (1992), 'Symposium on 'failed innovations'. Introduction', *Social Studies of Science*, **22**, 213-230.

Bresnahan, T. and M. Trajtenberg (1995), 'General Purpose Technologies: 'Engines of Growth'?', *Journal of Econometrics*, **65**(1), 83-108.

Brock D.C. (2014), 'Dudley Buck and the Computer that never was', *IEEE Spectr.*, April, 55-69.

Brock D.C. (2016), 'Will the NSA finally build its superconducting spy computer?', http://spectrum.ieee.org/computing/hardware/will-the-nsa-finally-build-its-superconducting-spy-computer, 24 Feb. 2016 [site accessed 4th Nov. 2016].

Brooks, H. (1994), 'The relationship between science and technology', *Research Policy*, **23**(5), 477-486.

Bruner, J. (1985), 'Narrative and paradigmatic modes of thought', in E. Eisner (ed.), *Learning and Teaching the Ways of Knowing*. University of Chicago Press: Chicago, 97-115.

Buck, D.A. (1956), 'The Cryotron – A Superconductive Computer Component', *Proceedings of the IRE* (Institute of Radio Engineers), **44**, 482-493.

Bush, V. (1945). *Science, the Endless Frontier: A Report to the President*. US Government Printing Office.

Callon, S. (1995), *Divided Sun. MITI and the Breakdown of Japanese High-Tech Industrial Policy 1975-1993*. Stanford University Press: Stanford.

Carr, D. (1986), 'Narrative and the real world: An argument for continuity', *History and Theory*, **25**(2), 117-131.

Ceruzzi, P.E. (2003), *A History of Modern Computing*, second edition [first edition 1998], MIT Press; Cambridge, Mass.

Constant, E. W. (1973), 'A model for technological change applied to the turbojet revolution', *Technology and Culture*, **14**(4), 553-572.

Constant, E.W. (1980), *The Origins of the Turbojet Revolution*. Johns Hopkins University Press: Baltimore.

Crane, H.D. (1962), 'Neuristor – a novel device and system concept', *Proc. IRE*, **50**, 2048-2060.

Crowe, J.W. (1957), 'Trapped-Flux Superconducting Films', *IBM Journal of Research and Development*, **1**, 294-303.

Cull, P. (2006), 'Caianiello and Neural Nets', in S. Termini (ed.), *Imagination and Rigor*, http://dx.doi.org/10.1007/88-470-0472-1_5, Springer: Milan.

Davidson A. (2015), *Personal memoir of a contributor to the IBM Josephson Computer Project*, Meeting held at the Department of Physics, University of Salerno, Italy, 15th September.

DeBresson, C. (1995), 'Predicting the most likely diffusion sequence of a new technology through the economy: The case of superconductivity', *Research Policy*, **24**(5), 685-705.

De Liso N. and G. Filatrella (2008), 'On technology competition: A formal analysis of the sailing-ship effect', *Economics of Innovation and New Technology*, **17**(5-6), 593-610.

de Solla Price, D. J. (1965), 'Is technology historically independent of science? A study in statistical historiography', *Technology and Culture*, **6**(4), 553-568.

Dorojevets, M., D. Strukow, A. Silver, A. Kleinsasser, F. Bedard, P. Bunyk, Q. Herr, G. Kerber and L. Abelson (2004), 'On the road towards superconductor computers: Twenty years later', in S. Luryi, J. Xu and A. Zaslavsky (eds), *Future Trends in Microelectronics*. Wiley: New York.

Dosi, G. (1982), 'Technological paradigms and technological trajectories: a suggested interpretation of the determinants and directions of technical change', *Research Policy*, **11**(3), 147-162.

Dray, W. H. (1971), 'On the nature and role of narrative in historiography', *History and Theory*, **10**(2), 153-171.

Economist, The (2016), The end of Moore's Law, in *The Economist Technology Quarterly*, issue 12th-18th March, Vol. 418, n. 8980.

Fisher, F.M., J.J McGowan and J.E Greenwood (1983), *Folded, Spindled and Mutilated. Economic Analysis and U.S. vs. IBM*. MIT Press: Cambridge, Mass.

Fischhoff, B. and R. Beyth (1975), 'I knew it would happen: Remembered probabilities of once—future things', *Organizational Behavior and Human Performance*, **13**(1), 1-16.

Flamm, K. (1987), *Targeting the Computer. Government Support and International Competition*. Brookings Institution: Washington D.C.

Flamm, K. (1988), *Creating the Computer. Government, Industry, and High Technology*. Brookings Institution: Washington D.C.

Foray, D. and M. Gibbons (1996), 'Discovery in the context of application', *Technological Forecasting and Social Change*, **53**(3), 263-277.

Freeman, C. (1973), 'A study of success and failure in industrial innovation', in B.R. Williams (ed.) *Science and Technology in Economic Growth*. Palgrave Macmillan: London, 227-255.

Freeman, C., (1982), *The Economics of Industrial Innovation*. University of Illinois at Urbana-Champaign's Academy for Entrepreneurial Leadership Historical Research Reference in Entrepreneurship. Available at SSRN: https://ssrn.com/abstract=1496190.

Freeman, C. and L. Soete (1997), *The Economics of Industrial Innovation*, Third edition, Pinter: London.

Fujimaki, A. (2012), 'Advancement of superconducting digital electronics', *IEICE Electronic Express*, **8**, 1720-1734.

Gallagher, W.J., E.P. Harris and M.B. Ketchen (2012), 'Superconducting at IBM – a centennial Review: Part I – Superconducting Computer and Device Application', *IEEE/CSC & ESAS European Superconductivity News Forum*, No. 21, July.

Garwin, R.L. (1957), 'An Analysis of the Operation of a Persistent-Supercurrent Memory Cell', *IBM Journal of Research and Development*, **1**, 304-308.

Gavroglu, K. (ed.) (2014), *History of Artificial Cold, Scientific Technological and Cultural Issues*. Springer: Dordrecht.

Goldstine, H.H. (1972), *The Computer from Pascal to von Neumann*. Princeton University Press: Princeton.

Goldstine, H.H. (1993), *The Computer from Pascal to von Neumann*. [second edition] Princeton University Press: Princeton.

Greenhalgh, T. and R. Peacock (2005), 'Effectiveness and efficiency of search methods in systematic reviews of complex evidence: audit of primary sources', *Bmj*, **331**(7524), 1064-1065, doi: https://doi.org/10.1136/bmj.38636.593461.68.

Hacklin, F., C. Marxt and F. Fahrni (2009), 'Coevolutionary cycles of convergence: An extrapolation from the ICT industry', *Technological Forecasting & Social Change*, **76**(6), 723-736.

Hypres (2015), Hypres to Develop Cryogenic Memory Solution for IARPA Superconducting Computers Program, Internet document http://www.hypres.com/

Jalali, S. and C. Wohlin (2012), 'Systematic literature studies: database searches vs. backward snowballing', in *Proceedings of the ACM-IEEE International Symposium on Empirical Software Engineering and Measurement*, September, ACM, 29-38, DOI: 10.1145/2372251.2372257.

Joas, C. and G. Waysand (2014), 'Superconductivity – A challenge to modern physics', chapter 5 in K. Gavroglu (ed.), *History of Artificial Cold, Scientific, Technological and Cultural Issues*. Springer: Dordrecht.

Johnston R.D. (1972), 'The internal structure of technology', in *The Sociological Review Monograph 18*, edited by P. Halmos, University of Keele, September, 117-130.

Josephson, B.J. (1962), 'Possible New Effects in Superconductive Tunnelling', *Physics Letters*, **1**, 251-253.

Josephson, B.J. (1973), 'The Discovery of Tunnelling Supercurrents', Nobel Lecture, 12[th] December.

Kahneman, D. (2011), *Thinking, Fast and Slow*. Farrar, Straus and Giroux: New York.

Kamin, K. A. and J.J. Rachlinski (1995), 'Ex post ≠ ex ante: Determining liability in hindsight', *Law and Human Behavior*, **19**(1), 89-104.

Kim, W., S.K. Han, K.J. Oh, T.Y. Kim, H. Ahn and C. Song (2010), 'The dual analytic hierarchy process to prioritize emerging technologies', *Technological Forecasting & Social Change*, **77**(4), 566-577.

Kotani, S., T. Imamura and S. Hasuo (1990), 'A Subnanosecond clock-Josephson 4-bit processor', *IEEE Journal of Solid-State Circuits*, **25**, 117-124.

Kranzberg, M. (1968), 'The disunity of science-technology', *American Scientist*, **56**(1), 21-34.

Krauss, L. M. and G.D. Starkmann (2004), 'Universal limits on computation', http://arxiv.org/abs/astro-ph / 0404510.

Kuhn, T. S. (1970), *The Structure of Scientific Revolutions*, [second edition, 1[st] edition 1962] University of Chicago Press: Chicago.

Kurtz, J.A. (1957), 'Superconducting Connection to Films', *IBM Journal of Research and Development*, **1**, 373.

Leoncini R. (2016), 'Learning-by-failing. An empirical exercise on CIS data', *Research Policy*, **45**(2), 376-386.

Likharev, K.K. (2012), 'Superconducting Digital Electronics', *Physica C: Superconductivity*, **482**, 6-18.

Logue, J. C. (1998), 'From vacuum tubes to very large scale integration: a personal memoir', *IEEE Annals of the History of Computing*, **20**(3), 55-68.

Malozemoff, A.P., W.J. Gallagher, R.L. Greene, R.B. Laibowitz and C.C. Tsuei (2012), 'Superconductivity at IBM – a Centennial Review: Part II – Materials and Physics', *IEEE/CSC & ESAS European Superconductivity News Forum*, No. 21, July.

Matricon, J. and G. Waysand (2003), *The Cold Wars. A History of Superconductivity*, [original French edition 1994], Rutgers University Press: New Brunswick, N.J.

Matisoo, J. (1966), 'Subnanosecond Pair-Tunneling to Single-Particle Tunneling Transitions in Josephson Junctions', *Applied Physics Letters*, **9**, 167-168.

Matisoo, J. (1980), 'The Superconducting Computer', *Scientific American*, **242**, 50-65.

Mazzucato, M. (2015). *The entrepreneurial state: Debunking public vs. private sector myths* (Vol. 1). Anthem Press.

Mendelsshon, K. (1977), *The Quest for Absolute Zero. The Meaning of Low Temperature Physics*, [second edition; first edition 1966], Taylor & Francis: London.

Merali, Z. (2012), 'Superconductor Breaks High-Temperature Record', *Nature*, 22 February, doi: 10.1038/nature.2012.10081.

Mina, A., R. Ramlogan, G. Tampubolon and J.S. Metcalfe (2007), 'Mapping evolutionary trajectories: Applications to the growth and transformation of medical knowledge', *Research Policy*, **36**(5), 789-806.

Mink, L. O. (1966), 'The autonomy of historical understanding', *History and Theory*, **5**(1), 24-47.

Mink, L. O. (1968), 'Change and causality in the history of ideas', *Eighteenth-Century Studies*, **2**(1), 7-25.

Mody C.C.M. (2017), *The Long Arm of Moore's Law. Microelectronics and American Science*, MIT Press: Cambridge, MA.

Mokyr, J. (1998), 'Induced technical innovation and medical history: an evolutionary approach', *Journal of Evolutionary Economics*, **8**(2), 119-137.

Mokyr, J. (2002), *The Gifts of Athena. Historical Origins of the Knowledge Economy*. Princeton University Press: Princeton.

Moore, G.E. (1965), 'Cramming More Components onto Integrated Circuits', *Electronics*, **38**, n. 8, 19 April, 114-117.

Mukhanov, O.A. (2014), 'Superconducting electronics: Solving energy problem in high-end computing', paper presented at the International Conference "Superconductivity for Energy", Paestum, Italy, 15-19 May.

Müller, K.A. and J.G. Bednorz (1987), 'The discovery of a class of high-temperature superconductors', *Science*, **237**, 1133-1139.

Narin, F., and E. Noma (1985), 'Is technology becoming science?', *Scientometrics*, **7**(3-6), 369-381.

Nelson, R.R. (2008), 'Factors affecting the power of technological paradigms', *Industrial and Corporate Change*, **17**(3), 485-497.

Nightingale, P. (2014), 'What is technology? Six definitions and two pathologies', Science Policy Research Unit (SPRU) Working Paper Series SWPS 2014-19, University of Essex http://dx.doi.org/10.2139/ssrn.2743113.

Nowotny, H. and U. Felt (1997), *After the Breakthrough: The Emergence of High-Temperature Superconductivity as a Research Field*. Cambridge University Press: Cambridge.

NRC – National Research Council (1995). *Evolving the High Performance Computing and Communications Initiative to Support the Nation's Information Infrastructure*, National Academy Press: Washington D.C.

NRC – National Research Council (1999). *Funding a Revolution. Government Support for Computing Research*, National Academy Press: Washington D.C.

ODNI – Office of the Director of National Intelligence (2014), IARPA Launches Program to Develop a Superconducting Computer, internet document: https://www.dni.gov/ index.php/ newsroom/ic-in-the-news/ic-in-the-news-2014/item/1146-iarpa-launches-program-to-develop-a-superconducting-computer.

Parmentier, R.D. (1967), *The Superconductive Tunnel Junction Neuristor*, unpublished PhD Thesis, University of Wisconsin.

Parmentier, R.D. (1969), 'Recoverable Neuristor Propagation on Superconductive Tunnel Junction Strip Lines', *Solid-State Electronics*, **12**, 287–297.

Platt, C. and D. Langford (1984), *Micromania: Whole Truth about Home Computers*, Gollanz: London.

Polanyi, M. (1956), 'Pure and applied science and their appropriate forms of organization', *Dialectica*, **10**(3), 231-242.

Polanyi, M. (1962), 'The republic of science: its political and economic theory', *Minerva*, **1**(1) 54-73.

Polkinghorne, D. E. (1995), 'Narrative configuration in qualitative analysis', *International Journal of Qualitative Studies in Education*, **8**(1), 5-23.

Popper, K. (1959), *The logic of scientific discovery*. (2005 edition), Routledge: London.

Reintjes, J.F. (1991), *Numerical Control. Making a New Technology*. Oxford University Press: New York.

Robinson, A.L. (1983), 'IBM Drops Superconducting Computer Project', *Science*, **222**, 492-494.

Rosenberg, N., and W.E. Steinmueller (2013), 'Engineering knowledge', *Industrial and Corporate Change*, **22**(5), 1129-1158.

Rothwell, R., C. Freeman, A. Horlsey, V.T.P. Jervis, A.B. Robertson and J. Townsend (1974), SAPPHO updated-project SAPPHO phase II, *Research Policy*, **3**(3), 258-291.

Rowell, J.M., M.R. Beasley and R.W. Ralston (1998), *Electronic Applications of Superconductivity in Japan*, WTEC Panel Report, International Technology Research Institute: Baltimore.

Schmalian, J. (2010), 'Failed Theories of Superconductivity', *Modern Physics Letters B*, **24**, 2679-2691.

Shachtman, T. (2000), *Absolute Zero and the Conquest of Cold*. Mariner Books: Boston.

Shin, S.W., G. Smith, J.A. Smolin and U. Vazirani (2014), 'How "Quantum" is the D-Wave Machine?', arXiv:1401.7087v1, 28 January.

Stajic, J. (2013), 'The Future of Quantum Information Processing', *Science*, **339**, 1163 – 1184, DOI: 10.1126/science.339.6124.1163.

Steinmueller, W. E. (1994), 'Basic research and industrial innovation', in M. Dodgson and R. Rothwell (eds), *The Handbook of Industrial Innovation*, Edward Elgar: Aldershot, 54-66.

Takagi, N., K. Marukami, A. Fujimaki, A. Yoshikawa, I. Koji and H. Honda (2008), 'Proposal of a desk-side supercomputer with reconfigurable data-paths using rapid single-flux-quantum circuits', *IEICE Transactions on Electronics*, **E91.C**, 350-355.

Tsukamoto, O. (2008), 'Overview of Superconductivity in Japan – Strategy Road Map and R&D Status', *Physica C: Superconductivity*, **468**, 1101-1111.

Velleman, J. D. (2003), 'Narrative explanation', *The Philosophical Review*, **112**(1), 1-25.

van Deft, D. (2014), 'The Cryogenic Laboratory of Heike Kamerlingh Onnes: An early Case of Big Science', chapter 4 in K. Gavroglu (ed.), *History of Artificial Cold, Scientific, Technological and Cultural Issues*. Springer: Dordrecht.

Vincenti, W. G. (1994), 'The retractable airplane landing gear and the Northrop "Anomaly": variation-selection and the shaping of technology', *Technology and Culture*, **35**(1), 1-33.

White, M. G. (1965), *Foundations of Historical Knowledge*. Harper & Row: New York.

Wilmoth, G.C. (2000), 'False-failed Innovation', *Joint Force Quarterly*, No. 23, March 2000, Autumn/Winter 1999-2000, 51-57.

Wohlin, C. (2014). Guidelines for snowballing in systematic literature studies and a replication in software engineering. In *Proceedings of the 18th International Conference on Evaluation and Assessment in Software Engineering*, Article N. 38, May, ACM.

Zelazny, R. (1976), *My Name is Legion*. Ballantine Books: New York.

Zimet, E., S. Starr, C. Lau and A. Ghosh (2009), 'Computer Science research Funding. How much is too Little?', Working Paper, Center for Technology and National Security Policy, National Defence University, Washington D.C.